\documentclass[12pt]{iopart}
\usepackage{amssymb,graphicx,iopams}

\def\be{\begin{equation}}
\def\ee{\end{equation}}
\def\bea{\begin{eqnarray}}
\def\eea{\end{eqnarray}}

\def\e{\mathrm{e}}

\def\E{{\cal E}}

\def\d{\mathrm{d}}

\def\exp{\mathrm{exp}}

\def\etal{\textit{et.al.}}
\def\ie{\textit{i.e.} }

\begin{document}

\title{The thermal denaturation of Peyrard-Bishop model with an external potential}

\author{A. Sulaiman$^{a,b,d}$, F.P. Zen$^{a,d}$, H. Alatas$^{c,d}$ and L.T.
Handoko$^{e,f}$}

\address{$^{a)}$Theoretical Physics Laboratory, THEPI Research
Division, Institut Teknologi Bandung, Jl. Ganesha 10, Bandung
40132, Indonesia.\\
$^{b)}$Badan Pengkajian dan Penerapan Teknologi, BPPT Bld.
II (19$^{\rm th}$ floor), Jl. M.H. Thamrin 8, Jakarta 10340,
Indonesia.\\
$^{c)}$Theoretical Physics Division,  Department of Physics, Bogor Agricultural University, Jl. Meranti, Kampus IPB Darmaga, Bogor 16680, Indonesia.\\
$^{d)}$Indonesia Center for Theoretical and Mathematical
Physics (ICTMP), Jl. Ganesha 10, Bandung 40132, Indonesia.\\
$^{e)}$Group for Theoretical and Computational Physics, Research
Center for Physics, Indonesian Institute of Sciences, Kawasan 
Puspiptek Serpong, Tangerang 15310, Indonesia.\\
$^{f)}$Department of Physics, University of Indonesia,
Kampus UI Depok, Depok 16424, Indonesia. }

\ead{albertus.sulaiman@bppt.go.id, handoko@teori.fisika.lipi.go.id}

\begin{abstract}
The impact of various types of external potentials to the Peyrard-Bishop DNA
denaturation is investigated through statistical mechanics approach. The
partition function is obtained using transfer integral method, and further the
stretching of hydrogen bond is calculated using  time independent perturbation
method. It is shown that all types of external potentials accelerate the
denaturation processes at lower temperature. In particular, it is argued that
the Gaussian potential with infinitesimal width reproduces a constant force at one end of DNA sequence as already done in some previous works.
\end{abstract}

\pacs{87.14.gk, 87.15.ad, 87.15.A-}

\maketitle

\section{Introduction}
\label{sec:intro}

It has been known that the base-pairs of double-helix break up and dissociate from 
each other to form  two separated random coils when a solution of DNA macromolecules 
is heated up to $80^\circ$C. This phenomenon is referred as the DNA denaturation or the 
thermal DNA melting (\cite{zhou} and references therein). On the other hand, at lower temperature before 
 melting two strands of DNA can also be separated by applying an oppositely directed force on  two 
strands at the DNA terminal point. This is known as the force-induced DNA melting \cite{zhou}. Theoretically, 
the phenomena have been discussed in many works, for instance the work by Hanke \etal  on the denaturation 
of stretched DNA \cite{hanke}. Also using the worm-like chain, the opened double-stranded DNA can be 
explained by a force exceeding certain critical value \cite{bhatta, orlandini}. On the other hand, 
the study of shear effect of pulling force shows that the shear unzipping of a heteropolymer would be 
similar to the unzipping in a tensile mode with sequence heterogeneity \cite{chakrabarti, pakas}.

It has been shown that a random external force would drastically change the phase diagram, while the ground 
state develops bubbles with various lengths as the random force fluctuation is increased.
The fluctuating force denaturates the DNA by a gradual increase of bubble sizes. This suggests the 
possibility of opening up local bubbles in the selective regions without breaking the whole DNA
\cite{kapri4}. The denaturation at lower temperature is realized by the enzymes, protein and so forth 
\cite{voet}. The effects are influenced by many factors such as the alkaline compound filling up 
a cell. Some works also investigated another effects, for instance the phase transition in a short DNA using 
the Peyrard-Bishop (PB) with additional delta function potential \cite{zhou2}. The paper argued that the 
denaturation is a localization-delocalization transition process. In the case without any external force, the 
denaturation is a second order transition, while under an external force it becomes the first-order one 
\cite{zhou}. Following the same line, the present paper deals with external potentials interacting to DNA within 
the PB model, and investigates its effects to the denaturation process.

This paper is organized as follows. First of all the PB model with
various external potentials is briefly introduced. It is then 
followed with the statistical mechanics formulation on the 
partition function of Hamiltonian under consideration. Before
ending the paper with summary, the detail calculation of hydrogen 
bond stretching using first order time independent perturbation 
and its thermal behavior are presented in Secs. \ref{sec:termo} 
and \ref{sec:behave}.

\section{The Model}
\label{sec:model}

Following the original PB model, the motion of DNA molecules is 
represented by transversal displacement, $u_n$ and $v_n$,
corresponding to the base displacement from their equilibrium
position along the direction of hydrogen bonds that is represented
by the Morse potential to connect two bases in a pair \cite{pbd1}.

The paper adopts the PB model with an additional external potential $V$ which has the same direction as 
Morse potential,
\bea
 H & = & \sum_n \frac{1}{2M} \dot{u}_n^2
  + \frac{\kappa}{2} \left( u_{n} - u_{n-1} \right)^2
  + \sum_n \frac{1}{2M} \dot{v}_n^2
 \nonumber\\
 & & + \frac{\kappa}{2} \left( v_{n} - v_{n-1} \right)^2
  + \frac{D}{2} \left( \e^{-\frac{\alpha}{2} \sqrt{2} \, (u_n-v_n)} - 1
\right)^2 + V(u_n-v_n) \; .
 \label{eq:pbasli}
\eea 
Performing a transformation to the center of mass
coordinate representing  the in-phase and out-phase transversal
motions, one can define $X_n = {(u_n+v_n)}/\sqrt{2}$ and $Y_n =
{(u_n-v_n)}/\sqrt{2}$. Then, it yields the PB Hamiltonian \cite{pbd1}, 
\bea
 H & = & \sum_n \frac{1}{2M} p_n^2
  + \frac{\kappa}{2} \left( X_{n+1} - X_{n} \right)^2
  + \sum_n \frac{1}{2M} P_n^2
 \nonumber\\
 & & + \frac{\kappa}{2} \left( Y_{n+1} - Y_{n} \right)^2
  + \frac{D}{2} \left( \e^{-\alpha/2  \, Y_n} - 1 \right)^2 + V(Y_n) \; ,
 \label{eq:pb}
\eea
where $D$ and $\alpha$ are the depth and inverse width of the potential
respectively. $p_n=M \dot{X_n}$, $P_n=M\dot{Y_n}$ and $\kappa$ is the spring
constant.

It should be remarked that some previous works suggested that the $n-$th
nucleotide in one strand might preferably interacts with the $(n \pm h)-$th
nucleotides in another strand with $h = 4$ or $5$ \cite{pbd2,pbd3,zts1}. However
for the sake of simplicity, throughout the paper let us consider the original PB
model with taking into account only the nearest neighbor nucleotides of another
strand. The extension to the cases of $h \ne 1$ should require extensive
numerical studies that is out of our current interest. The variable $X_n$ is
decoupled from $Y_n$ corresponding merely to the linear chain, while $Y_n$
represents the stretching motion that is our main interest. So, from now let us
ignore the $X_n$ part in further calculation.


\section{Mechanical statistics calculation}
\label{sec:termo}

Now let us calculate the mechanical statistics behavior of 
current system using transfer integral method. Having certain 
hamiltonian of an equilibrium system, one can extract some physical observables
through statistical mechanics approach. The approach is
particularly suitable to investigate the dynamics of DNA with
thermal fluctuations since the exact solution of its equation of motion never
exists \cite{peyrard}. Within the PB model, the DNA denaturation
and its melting temperature have previously been studied in term
of temperature \cite{pbd1,pbd3}. It was argued that the
denaturation can be induced by energy localization due to
nonlinear effects, and also should be highly influenced by
external interactions like $V_n$ in the present case.

Using the hamiltonian in Eq. (\ref{eq:pb}), one can consider
the partition function to further calculate some thermodynamics
variables. In \cite{pbd1,peyrard}, the calculation has been
performed using transfer integral method for $V = 0$. This paper
follows the same procedure. In the canonical ensemble the
partition function density is related to the Hamiltonian by $Z
\propto \exp(\beta H)$. For the present case it reads, 
\bea
  Z & = & \int \int \prod_{n=1}^N \prod_{n=0}^N \d P_n \d Y_n \,
  \exp \left\{ -\beta \, \left[ \sum_n \frac{P_n^2}{2M} \right.\right.
    + \frac{\kappa}{2} \left( Y_{n} - Y_{n-1} \right)^2 \nonumber\\
  & & \left.\left.
  + \frac{D}{2} \left( \e^{-\alpha/2 \, Y_n} - 1 \right)^2
    +V(Y_n) \right] \right\} \delta(Y_N -Y_0)\; ,
  \label{eq:partisi1}
\eea
where $\beta = 1/{k T}$ with $k$ is the Boltzmann constant. It can further be decomposed
into its momentum and coordinate spaces, $Z = Z_P Z_Y$ with,
\bea
    Z_P & = & \int  \prod_{n=1}^N \d P_n \, \exp \left( -\beta
      \sum_n \frac{P_n^2}{2M} \right) \; ,
    \label{eq:ZP1} \\
    Z_Y & = & \int \prod_{n=0}^N \d Y_n \, \delta(Y_N - Y_0)
      \exp \left\{-\beta \left[ \frac{\kappa}{2} (Y_{n}-Y_{n-1})^2
  \right.\right.
  \nonumber\\  && \left . \left.
  + \frac{D}{2} \left( \e^{-\alpha/2 \, Y_n} - 1 \right)^2 +
      V(Y_n) \right] \right\} \; ,
    \label{eq:ZY1}
\eea

The Gaussian integration in Eq. (\ref{eq:ZP1}) yields,
\be
  Z_P = \left( \sqrt{\frac{2\pi M }{\beta}} \right)^N \; .
  \label{eq:ZP2}
\ee
On the other hand, Eq. (\ref{eq:ZY1}) reads,
\be
  Z_Y = \int \prod_{n=0}^{N} \d Y_n \, \delta(Y_N-Y_0)
  \prod_{n=1}^{N} \e^{-\beta \Theta(Y_n,Y_{n-1})} \; ,
  \label{eq:ZY2}
\ee
where, 
\be
\Theta(Y_n,Y_{n-1})= \frac{\kappa}{2} (Y_{n}-Y_{n-1})^2 +
\frac{D}{2} (\e^{-\frac{\alpha}{2}Y_n}-1)^2 + V(Y_n) \; . 
\ee
This can be solved  using transfer integral method.

Shifting the operator $Y_{n-1} \rightarrow Y_n$ and defining an
eigenfunction $\phi(Y_n)$ that satisfies \cite{peyrard},
\be
\int \d Y_{n-1} \, \e^{-\beta \Theta(Y_n,Y_{n-1})}
\phi_i(Y_{n-1})= \e^{-\beta E_i} \phi_i(Y_n) \; ,
  \label{eq:ZY4}
\ee
Eq. (\ref{eq:ZY2}) becomes,
\be
  Z_Y= \int \prod_{n=0}^{N} \d Y_n \, \delta(Y_N-Y_0) \, \e^{-\beta
E_i}\phi_i(Y_n) \; .
  \label{eq:ZY5}
\ee
Since $\delta(Y_N-Y_0)=\sum_i \phi_i^*(Y_N)\phi_i(Y_0)$
and taking the normalization $\int \d Y \, \phi^* \phi =1$, one gets,
\be
  Z_Y= \sum_i \e^{-\beta N E_i} \; .
  \label{eq:ZY6} \\
\ee

Let us find the eigenfunction $\phi(y)$ and the eigenvalue $E$ of
Eq.({\ref{eq:ZY4}}). First of all at continuum limit, $Y_{n-1} \sim x$ and $Y_n
\sim y$, Eq. (\ref{eq:ZY4}) reads,
\be
  \int \d x \, \phi(x)
  \exp \left\{ -\beta \left[ \frac{\kappa}{2} (y - x)^2
  + \frac{D}{2} \left( \e^{-\alpha/2 \, y} - 1 \right)^2 + V(y)
  \right]\right\} = \e^{-\beta E} \, \phi(y) \; .
   \label{eq:TI1}
\ee
Now assuming that the harmonic term is dominant, $x = y + z$ can be expanded as
a Taylor series in the power of $z$. The expansion up to the second order
yields,
\bea
  && \int \d z \, \e^{-\frac{\beta \kappa}{2}z^2}\left[ \phi(y)+\frac{\d \phi}{\d y} z
          +\frac{1}{2}\frac{\d^2 \phi}{\d y^2} \right]
  \nonumber\\
  &&   = \exp \left\{ -\beta \left[ E
    - \frac{D}{2} \left( \e^{-\alpha/2 \, y} - 1 \right)^2
    + V(y) \right] \right\} \phi(y) \; .
   \label{eq:TI2}
\eea
Again, this is just the Gaussian integration and the result is \cite{feynman},
\bea
   \phi(y) + \frac{1}{2 \beta \kappa} \frac{\d^2 \phi}{\d y^2}
          & = & \exp \left\{ -\beta \left[ E
        + \frac{1}{2 \beta} \ln \left(\frac{2\pi}{\beta \kappa} \right)
    \right.\right.
    \nonumber\\
  && \left.\left. \; \; \; \; \; \; \;
        - \frac{D}{2} \left( \e^{-\frac{\alpha}{2}y} - 1 \right)^2
        + V(y) \right] \right\} \phi(y)
   \nonumber\\
   & \approx &
          \left[ 1 - \beta E
      - \frac{1}{2} \ln \left( \frac{2\pi}{\beta\kappa} \right)
      \right]\phi(y)
   \nonumber\\
   && + \beta \left[ \frac{D}{2} \left( \e^{-\alpha/2 \, y} - 1 \right)^2
      + V(y) \right]\phi(y) \; .
   \label{eq:TI5}
\eea 
This is nothing else than the Schr$\ddot{\mathrm{o}}$dinger like equation,
\be
    -\frac{1}{2 m_0} \frac{\d^2 \phi}{\d y^2} + \frac{D}{2} \left[ \e^{ -\alpha/2 \, y} -1 \right]^2 \phi
    +  V(y) \phi = \bar{E} \phi \; ,
    \label{eq:TI6}
\ee 
where $m_0 = \kappa \beta^2$ and $\bar{E} = E + 1/{(2\beta)}
\ln \left[ 2\pi/(\beta \kappa) \right]$. The problem is therefore
turned into finding the eigenvalue of Eq. (\ref{eq:TI6}).

Eq. (\ref{eq:TI6}) then yields an eigen equation $H \phi_m= \bar{E}_m \phi_m$ with Hamiltonian, 
\be
  H = \frac{1}{2 m_0} \frac{\d^2 }{\d y^2} + \frac{D}{2} \left[ \e^{-\alpha/2 \, y} - 1 \right]^2+ V(y) \, ,
  \label{eq:TI7}
\ee
and $m$ is an integer. For a constant force $V_0$, $V(y) = V_0 y$, the transformation 
$\phi=\exp(\beta V_0 y)\psi$ reproduces the Schr$\ddot{\mathrm{o}}$dinger equation with 
Morse potential \cite{singh}. Unfortunately, the equation can be solved only 
for few special cases. Generally, the solution can be obtained perturbatively by assuming that 
the potential $V(y)$ is small enough compared to the Morse potential. Under this assumption, 
the eigen equation can be solved using time independent perturbation theory in standard 
quantum mechanics. However, it should be emphasized that the problem remains a classical one, 
and the quantum mechanics is borrowed only for technical reason since the calculation yields the 
 Schr$\ddot{\mathrm{o}}$dinger equation.

In quantum mechanics the perturbation is applied by expanding the Hamiltonian ($H$) 
in a form of $H = H_0 + \epsilon V$ with $H_0$ is the hamiltonian in Eq. (\ref{eq:TI7}) without the 
potential, while $\epsilon$ is a small parameter. According to the time-independent 
perturbation theory, the eigenvalue and its eigenfunction can be written as \cite{mueller},
\bea
   \bar{E}_m & = & \bar{E}_m^{(0)} + \epsilon \bar{E}_m^{(1)} + \epsilon^2 \bar{E}_m^{(2)} +...\; , \\
   \phi_m(y) & = & \phi_m^{(0)}(y) + \epsilon \phi_m^{(1)}(y)+\epsilon^2 \phi_m^{(2)}(y)+... \; .
    \label{eq:perturb1}
\eea
The eigenvalue for $H_0$ is determined by $H_0\phi^{(0)}=\bar{E}_0^{(0)}\phi^{(0)}$, that is,
\be
  -\frac{1}{2 m_0} \frac{\d^2 \phi^{(0)}}{\d y^2}
  + \frac{D}{2} \left( \e^{-\alpha/2 \, y} - 1 \right)^2 \phi^{(0)}
  = \bar{E}^{(0)} \phi^{(0)} \; ,
  \label{eq:TI8}
\ee
where $\bar{E}^{(0)} = E^{(0)} + 1/(2 \beta)\ln [ 2\pi/(\beta\kappa)]$.
 The equation is just the well known Schr$\ddot{\mathrm{o}}$dinger equation with Morse potential 
and the solution has been derived in \cite{peyrard}. Substituting $y'=(1/2) \alpha y$, $\lambda = 2\sqrt{m_0
D}/\alpha$, $\E^{(0)} = {8 m_0}/{\alpha^2\bar{E}^{(0)}}$ and $z =
2 \lambda \, \exp ({-y'})$, one arrives at the eigenvalue problem
for ${\beta \sqrt{\kappa D}}/\alpha > 1/2$ as
\cite{sethi,mueller},
\be
    -\frac{\d^2 \phi^{(0)}}{\d y'^2} +\lambda^2\left(\e^{-2y'}-2\e^{-y'}\right)
    \phi_{m}^{(0)}= \E_{m}^{(0)} \phi_{m}^{(0)} \; .
    \label{eq:TI9}
\ee
The eigenvalue and its eigenfunction are,
\bea
  \E_{m}^{(0)} & = &
    - \left( \lambda-{m}-\frac{1}{2} \right)^2 \; , \\
  \label{eq:TI10}
  \phi_{m}^{(0)} & = &
    N_{m} z^{b_{m}} \e^{-z/2} \, L_{m}^{2b_{m}}(z) \; ,
  \label{eq:TI11}
\eea 
where $b_{m} = \lambda - {m} - 1/2$ and $N_{m} = {m}! \left[
\Gamma \left({m} + 1 \right) \Gamma \left( 2\lambda - {m} \right)
\right]^{-1/2}$ with $\Gamma$ is the Gamma function.
$L_{m}^{2b_{m}}(z)=(z^{-2b_{m}}e^z/m!)(d^{m}(e^{-z
}z^{m+2b_{m}})/dz^{m})$ is the associated Laguerre polynomial or
generalized Laguerre polynomials.

Based on these transformation, using the standard
procedure in quantum mechanics one obtains the solution up to the first order \cite{mueller}, 
\bea
   \E_m & = & \E_m^{(0)} + \epsilon \E_m^{(1)} \; , \\
   \phi_m(y') & = & \phi_m^{(0)}(y') + \epsilon \phi_m^{(1)}(y') \; ,
    \label{eq:perturb3}
\eea
where,
\bea
   \E_m^{(1)}&=& \int \d y' \, \phi_m^{(0)\ast}(y')V(y') \phi_m^{(0)}(y') \; , \\
   \phi_m^{(1)}(y') &=& \sum_{k\neq m} \frac{\int \d y' \, \phi_k^{(0)\ast}V(y')
   \phi_m^{(0)}(y')}{\E_m^{(0)}-\E_k^{(0)}}\phi_k^{(0)}(y') \; .
  \label{eq:perturb2}
\eea

Finally the partition function of  $y$ is given by,
\be
  Z_y =  \sum_m \e^{-\beta N E_m} \; ,
  \label{eq:partisY}
\ee
with the eigenvalue,
\be
  E_m = \frac{\alpha^2}{8 \kappa \beta^2} \E_m
  + \frac{1}{2\beta} \ln \left( \frac{\beta \kappa}{2\pi} \right) \; .
  \label{eq:eigenY}
\ee

\section{Thermal denaturation}
\label{sec:behave}

One of relevant parameters in the study of DNA denaturation is the
mean stretching, $\langle y_n \rangle$, of the hydrogen bond. It is defined as,
\be
  \langle y_n\rangle  = \frac{1}{Z} \int \d y_n \d p_n \, \prod_{n=1}^N y_n \e^{-\beta H} \; .
  \label{eq:hidrogen1}
\ee
The average value of hydrogen bond stretching can be calculated through,
\be
  \langle y\rangle = \frac{\sum_{i=1}^N \langle \phi_i(y)\mid y \mid
\phi_i(y)\rangle  \e^{-N \beta \bar{E}_i}}
  {\sum_{i=1}^N \langle \phi_i(y)\mid  \phi_i(y)\rangle  \e^{-N \beta \bar{E}_i}}  \; .
  \label{eq:hidrogen2}
\ee

At the limit of large $N$, $\langle y\rangle$ is dominated by the ground state
\cite{peyrard}, that is $\langle y\rangle = \int \d y (\phi_0(y))^* y
\phi_0(y)$. Up to the leading term of perturbation, \ie
$\phi_0(y) = \phi_0^{(0)}(y) + \epsilon \phi_0^{(1)}(y)$, Eq.
({\ref{eq:hidrogen2}}) reads,
\bea
  \langle y\rangle & = & \int \d y \, \phi_0^{(0)*}(y) y \phi_0^{(0)}(y)
  \nonumber\\
  && + \epsilon \left[
  \int \d y \, \phi_0^{(0)*}(y) y \phi_0^{(1)}(y)
  + \int \d y \, \phi_0^{(1)*}(y) y \phi_0^{(0)}(y) \right] \; .
  \label{eq:hidrogen4}
\eea
The higher order of eigenvalue and eigenfunction are given by,
\be
   \E_0^{(1)} =  \int \d y  \phi_n^{(0)\ast}(y)V(y) \phi_n^{(0)}(y) \; ,
   \label{eq:satu1}
\ee
and,
\bea
 \phi_0^{(1)}(y) &=&  \sum_{k\neq 0} \frac{\int \d y \, \phi_k^{(0)\ast}V(y)y
 \phi_0^{(0)}(y)}{\E_0^{(0)}-\E_k^{(0)}}\phi_k^{(0)}(y)\;,
 \nonumber\\
  &=& \frac{\int \d y \, \phi_1^{(0)\ast} V(y)
 \phi_0^{(0)}}{\E_0^{(0)}-\E_1^{(0)}}\phi_1^{(0)}(y) +\frac{\int \d y \, \phi_2^{(0)\ast}
 V(y)
 \phi_0^{(0)}}{\E_0^{(0)}-\E_2^{(0)}}\phi_2^{(0)}(y) + \cdots \; .
  \label{eq:satu2}
\eea
Then Eqs. (\ref{eq:TI10})$\sim$(\ref{eq:perturb2}) yield,
\bea
  \langle y\rangle &=& I_{(00)} + \epsilon
  \left[ \frac{I_{(10)}}{\E_0^{(0)} - \E_1^{(0)}} \int \d y \, \phi_0^{(0)*}(y) y
   \phi_1^{(0)}(y) \right. \nonumber\\
  & & \left. + \frac{ I_{(10)}}{\E_0^{(0)} - \E_1^{(0)}} \int \d y \, \phi_1^{(0)*}(y) y \phi_0^{(0)}(y)\right. \nonumber\\
   && \left. + \frac{ I_{(20)}}{\E_0^{(0)} - \E_2^{(0)}} \int \d y \,
\phi_0^{(0)*}(y) y \phi_2^{(0)}(y) \right. \nonumber\\
   & & \left. + \frac{ I_{(20)}}{\E_0^{(0)} - \E_2^{(0)}} \int \d y \,
\phi_2^{(0)*}(y) y \phi_0^{(0)}(y) + \cdots \right] + O(\epsilon^2)  \; ,
     \label{eq:hidrogen5}
\eea
where,
\bea
  I_{00} &=& \int \d y \,  \phi_0^{(0)*}(y)V(y)  \phi_0^{(0)}(y) \; ,  \\
  I_{10} &=& \int \d y \,  \phi_1^{(0)*}(y)V(y) \phi_0^{(0)}(y) \; , \\
  I_{20} &=& \int \d y \,  \phi_2^{(0)*}(y)V(y)  \phi_0^{(0)}(y) \; ,
  \label{eq:integralnya}
\eea
while the eigenvalues and eigenfunctions are given by,
\bea
   \E_0^{(0)} &=&-\left(\lambda-\frac{1}{2}\right)^2 \;    ,\\
   \E_1^{(0)} &=&-\left(\lambda-\frac{3}{2}\right)^2  \;    ,\\
   \E_2^{(0)} &=&-\left(\lambda-\frac{5}{2}\right)^2  \;    .\\
   \cdots  \nonumber
   \label{eq:eigenval}
\eea
Here,
\bea
  \phi_0^{(0)} & = &
  \frac{(2\lambda)^{\lambda - 1/2}}{\sqrt{\Gamma(2\lambda)}} \,
    \e^{-\lambda  \, \exp ( -\alpha/2 \, y )} \,
      \e^{-\alpha/2 (\lambda - 1/2) y} \; , \\
  \phi_1^{(0)} & = &\frac{2(2\lambda)^{\lambda -
\frac{3}{2}}}{\sqrt{\Gamma{(2)} \Gamma{(2\lambda-1)}}} \,
    \e^{-\lambda \, \exp ( -\alpha/2 \, y )} \,
      \e^{-\alpha/2 (\lambda - 3/2) y}\nonumber\\
      && \times \left[ -\lambda \e^{-\alpha/2 \, y} + (\lambda - 1) \right] \; ,
\\
      \phi_2^{(0)} &
= & \frac{4(2\lambda)^{\lambda - 5/2}}{\sqrt{\Gamma{(3)}
\Gamma{(2\lambda-2)}}} \,
    \e^{-\lambda \, \exp ( -\alpha/2 \, y )} \,
      \e^{-\alpha/2 (\lambda - 5/2) y} \nonumber \\
      && \times \left[ \lambda^2 \, \e^{-\alpha y} - \lambda
(2 \lambda - 3) \, \e^{-\alpha/2 \, y}
      +\frac{1}{4}( 2 \lambda - 3)( 2 \lambda - 4) \right] \; ,
      \label{eq:eigenfuncs}
\eea 
where $\lambda= 2\sqrt{\kappa D}/(\alpha k_B T)$.

\begin{figure}[t]
 \centering
 \includegraphics[width=14cm]{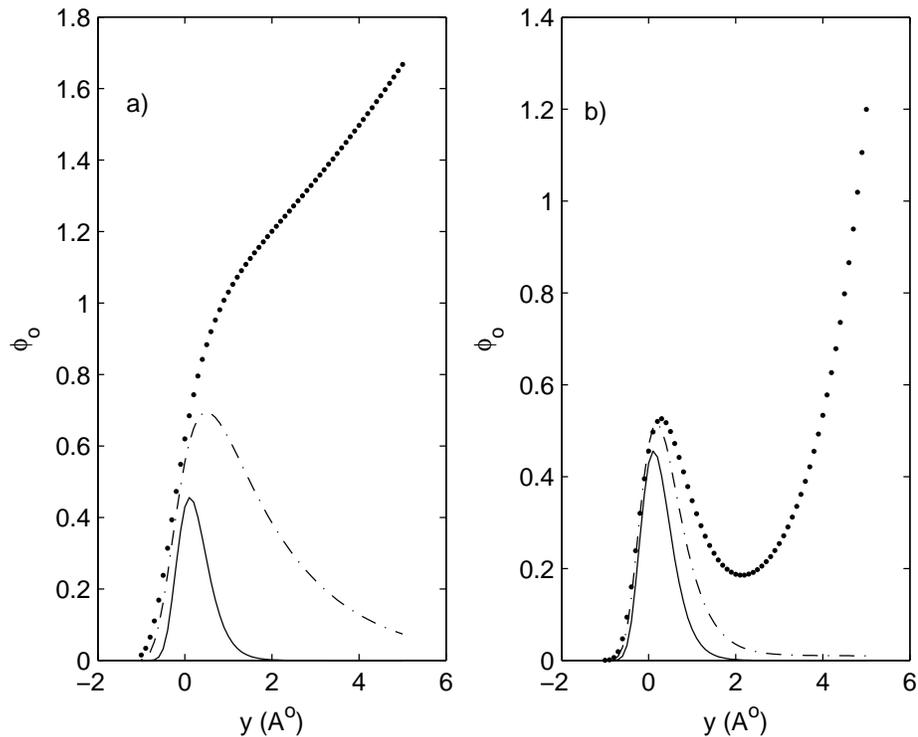}
 \caption{The eigenfunction without (left) and with (right) external linear potential as a
function of the displacement in various temperatures : 100K (solid line),
300K (dashed-line) and 500K (dotted-line) in the left; and 100K (solid
line), 150K (dashed-line) and 200K (dotted-line) in the right figures.}
 \label{fig:wave}
\end{figure}

Considering the case of linear potential $V(y) = V_0 y$, the wave function is given by,
\be
 \phi_0(y) = \phi_0^{(0)}+  \epsilon \left[ \frac{V_0\int \d y \, \phi_1^{(0)\ast}y
 \phi_0^{(0)}}{\E_0^{(0)}-\E_1^{(0)}}\phi_1^{(0)}(y) +  \frac{V_0\int \d y \, \phi_2^{(0)\ast} y
 \phi_0^{(0)}}{\E_0^{(0)}-\E_2^{(0)}}\phi_2^{(0)}(y) + \cdots \right] \; .
  \label{eq:wave1}
\ee 
The variation of wave function in respect to temperature is
depicted in Fig. {\ref{fig:wave}}  for the same parameters as
\cite{pbd3}, that is $D = 0.04$ eV, $\alpha=4.45$ {\AA}$^{-1}$,
$\epsilon = 0.01$, $\kappa=0.06$ eV/{{\AA}$^2$} and $V_0 =
15$ pN.

The ground state of eigenfunction for linear potential gives the
weighting factor for the calculation of DNA displacement $\langle
y \rangle$ \cite{pbd1}. The figure shows that at low temperature,
the peak is close to the minimum of Morse potential and decays
exponentially as the displacement increases. Without the external
potential, the eigenfunction is increasing drastically at the melting
temperature, that is about $500$K as already pointed out in
\cite{pbd3}. On the other hand, the same behavior occurs at lower
temperature when the external potential is applied to the system, \ie
at about $200$K.

\begin{figure}[t]
 \centering
 \includegraphics[width=14cm]{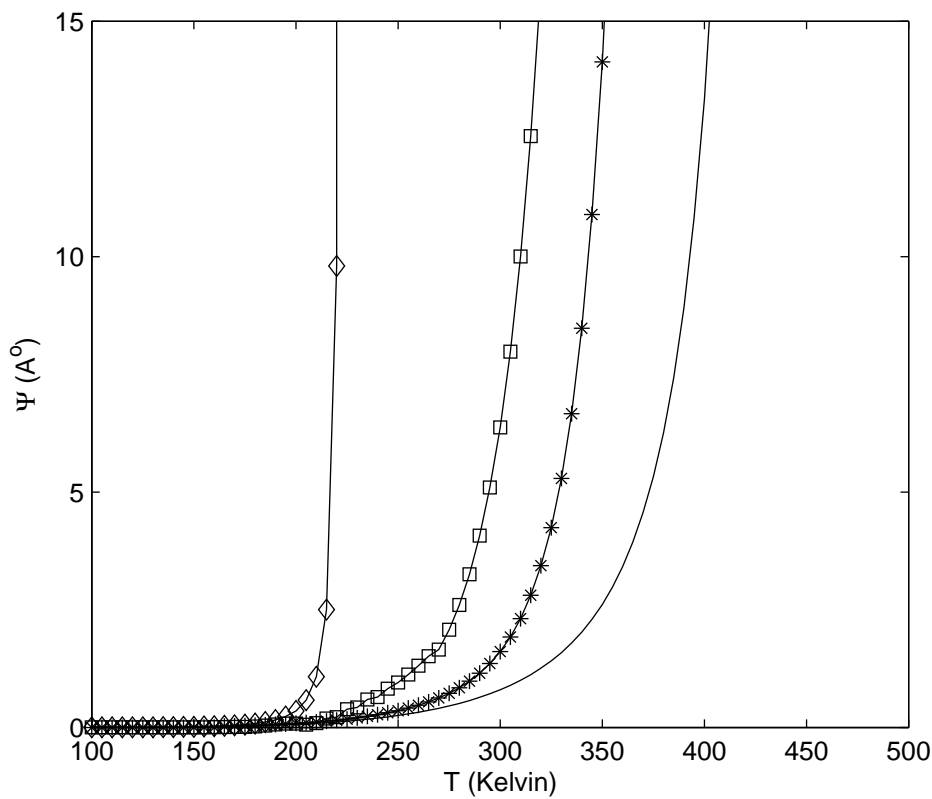}
 \caption{The hydrogen bond stretching as a function of melting temperature
using the same parameters as \cite{pbd3} for various potentials :
$V(y) = 0$ (plain line), $V(y) = V_0 y$ (line with diamonds),
$V(y) = V_0 \sin(2.5 y)$ (line with squares) and  $V(y) = V_0 \e^{-10 y^2}$ (line with stars) .}
 \label{fig:denat}
\end{figure}

From Eq. (\ref{eq:hidrogen5}), the influence of various external potentials into
denaturation processes is depicted in Fig. {\ref{fig:denat}} using the 
same parameters as \cite{pbd3}. The figure describes the 
effects of linear potential $V(y) = V_0 y$, sinusoidal potential $ V(y) 
= V_0 \sin(a y)$ and Gaussian potential $V(y) = V_0 \e^{-a y^2}$ with $a$ is a
constant. It should be noted that the result coincides to the
previous work for zero external potential \cite{pbd3}. The figure
shows that the critical temperature is approximately around $400$K
without external potential, and around $200$K with constant external
potential. Nevertheless the critical temperature becomes around $350$K
for periodic sinusoidal force with $a=2.5$. The same behavior of
$\langle y \rangle$ for non forcing condition has also been
obtained in \cite{sung} using flexible chain model with typical
parameter $D = 0.25eV$, $\alpha = 2.8$ {\AA}$^{-1}$ and numerical
calculation of the stochastic description of flexible polymer
models.

\section{Summary}

The effect of external potential to the Peyrard-Bishop DNA
denaturation has been investigated for various types of potential
: linear potential $ V(y) = V_0 y$, sinusoidal potential $V(y) = V_0 \sin(a y)$ and Gaussian potential 
$V(y) = V_0 \e^{-a y^2}$ with $a$ is a constant. The DNA denaturation and its melting temperature
have further been studied in the framework of statistical
mechanics approach. The calculation of partition function has been
performed perturbatively using transfer integral method to obtain
analytically the mean stretching of hydrogen bond.

The present paper support the results of previous works. In a cell, DNA strands can 
be separated by applying certain external potential \cite{rouzina1}, or in chemical terms by enzymes
whose interactions with DNA could make the strand separation 
thermodynamically favorable at ambient temperature \cite{sasaki}.
It has been shown that two strands of double-stranded DNA can be
separated  by applying $V_0 \sim 15$ pN force at room temperature. 
The model also predicts that the DNA overstretching force should be a decreasing 
function  of temperature, or in another word the melting temperature should be a decreasing
function of the applied force \cite{rouzina1, rouzina2}.

\section*{Acknowledgments}

AS thanks the Group for Theoretical and Computational Physics LIPI for warm
hospitality during the work. This work is funded by the Indonesia Ministry of
Research and Technology and the Riset Kompetitif LIPI in fiscal year 2011 under
Contract no.  11.04/SK/KPPI/II/2011. FPZ thanks Research KK ITB 2012 and Hibah
Kompetensi Kemendikbud RI 2012.

\section*{References}

\bibliographystyle{elsarticle-num}
\bibliography{Sulaiman}

\end{document}